\documentclass[11pt,letterpaper,twoside]{article}
\usepackage[english]{babel}
\usepackage{cite}
\usepackage{xcolor}
\usepackage{graphicx,epsfig,subfigure}
\usepackage{amsmath}
\usepackage{amssymb,latexsym}
\usepackage[all,2cell]{xy}
\UseTwocells
\usepackage{enumitem}
\usepackage{listings}
\usepackage{ctable}
\usepackage{lscape}
\usepackage[font=footnotesize]{caption}
\usepackage{array}
\usepackage{tabulary}
\usepackage{multirow}
\usepackage[cp1250]{inputenc}
\usepackage{mdframed}
\usepackage{subfiles}
\usepackage{url}

\usepackage{setspace}

\usepackage[referable]{threeparttablex}
\usepackage{multirow}
\usepackage{tabu}
\usepackage{enumitem}

\usepackage{algorithm,algorithmicx}
\usepackage{algpseudocode}
\usepackage{multicol}

\usepackage{fancyhdr}
\newcommand{\ace}{\textsf{ACE}}
\newcommand{\rc}{\textit{rc}}
\renewcommand{\sc}{\textit{sc}}

\newcommand{\aue}{$\mathcal{AE}$}
\newcommand{\hash}{$\mathcal{H}$}

\newcommand{\aceae}{\ace-\aue\textsf{-128}}
\newcommand{\acehash}{\ace-\hash\textsf{-256}}
\newcommand{\aceenc}{\ace-$\mathcal{E}$}
\newcommand{\acedec}{\ace-$\mathcal{D}$}

\newcommand{\simeck}{$\mathtt{SB\text{-}64}$}
\newcommand{\simecki}{$\mathtt{SB\text{-}64}_1$}

\newcommand{\lfsrc}{{\tt lfsr\_c}}

\newcommand{\ie}{\emph{i.e.}}
\newcommand{\eg}{\emph{e.g.}}
\newcommand{\etc}{\emph{etc.}}

\newcommand{\wage}{{\textsf{WAGE}}}
\newcommand{\wg}{{\textsf{WG}}}
\newcommand{\wagename}{{\textsf{W}$\mathcal{A}$\textsf{G}$\mathcal{E}$}}
\newcommand{\wgp}{{\textsf{WGP}}}

\newcommand{\sbx}{\textsf{SB}}
\newcommand{\lfsr}{{\textsf{LFSR}}}

\newcommand{\FF}{$\mathbb{F}_{2^7}$}

\newcommand{\wageenc}{\textsf{WAGE-$\mathcal{E}$}}
\newcommand{\wagedec}{\textsf{WAGE-$\mathcal{D}$}}

\newcommand{\wageae}{\wage-$\mathcal{AE}$\textsf{-128}}

\newcommand{\wagelfsr}{{\tt wage\_lfsr}}
\newcommand{\wgphw}{{\tt WGP}}
\newcommand{\sbxhw}{{\tt SB}}

\makeatletter

\renewcommand\section{\@startsection {section}{1}{\z@}%
                                   {-0.5ex \@plus -2ex \@minus -.2ex}%
                                   {0.2ex \@plus.2ex}%
                                   {\normalfont\Large\bfseries}}
\renewcommand\subsection{\@startsection{subsection}{2}{\z@}%
                                     {-0.5ex\@plus -1ex \@minus -.2ex}%
                                     {0.2ex \@plus .2ex}%
                                     {\normalfont\large\bfseries}}
\renewcommand\subsubsection{\@startsection{subsubsection}{3}{\z@}%
                                     {-0.5ex\@plus -1ex}%
                                     {0.2ex \@plus .2ex}%
                                     {\normalfont\normalsize\bfseries}}
\makeatother

\begin{document}
\thispagestyle{empty}

\begin{minipage}{\linewidth}
  \centering
  {\Large
   Hardware Design and Analysis of the {\ace} and {\wage} Ciphers
  }\vspace{1ex}

  Mark D.\@ Aagaard, Marat Sattarov, and Nu\v{s}a Zidari\v{c}\\
  Department of Electrical and Computer Engineering\\
  University of Waterloo, Ontario, Canada\\
  \texttt{\{maagaard,msattaro,nzidaric\}@uwaterloo.ca}

\end{minipage}

\renewcommand{\thefootnote}{}
\footnotetext{This work was supported in part by the Canadian National
  Science and Engineering Research Council (NSERC); the Canadian
  Microelectronics Corp (CMC); and
Grant 60NANB16D289 from the U.S. Department of Commerce,
National Institute of Standards and Technology (NIST).}

\vspace{-1ex}

\begin{abstract}
This paper presents the hardware design and analysis of {\ace} and \wage,
two candidate ciphers for the NIST Lightweight Cryptography
standardization.  Both ciphers use sLiSCP's unified sponge duplex
mode. {\ace} has an internal state of 320 bits, uses three 64\,bit
Simeck boxes, and implements both authenticated encryption and hashing.
{\wage} is based on the Welch-Gong stream cipher and provides
authenticated encryption. {\wage} has 259 bits of state, two 7\,bit
Welch-Gong permutations, and four lightweight 7\,bit S-boxes. {\ace} and
{\wage} have the same external interface and follow the same I/O protocol
to transition between phases.  The paper illustrates how the {\ace}
and {\wage} algorithms were translated into hardware and some of the
key design decisions in this process.  
The paper reports area, power, and energy results for both serial and
parallel (unrolled) implementations using four different ASIC
libraries: two 65\,nm libraries, a 90\,nm library, and a 130\,nm library.
{\ace} implementations range from a throughput of 0.5
bits-per-clock cycle (bpc) and an area of 4210\,GE (averaged across the
four ASIC libraries) up to 4\,bpc and 7260\,GE.  {\wage} results range
from 0.57\,bpc with 2920\,GE to 4.57\,bpc with 11080\,GE.
\end{abstract}

\vspace{-1ex}

\section{Introduction}
In 2013, NIST started the Lightweight Cryptography (LWC)
project~\cite{lwc}, with the end goal of creating a portfolio of
lightweight algorithms for authenticated encryption with associated
data (AEAD), and optionally hashing, in constrained environments
\cite{call}. For hardware-oriented lightweight algorithms, hardware
implementation results are an important criteria for assessment and
comparison.  In the first round of the LWC evaluation, more than half
of the candidates~\cite{lwcr1} reported hardware implementation
results or their estimates, ranging from complete implementation and
analysis to partial implementation results and theoretical estimates
based on gate count. Various amounts of analysis, such as area
reported only for a cryptographic primitive used or thorough area
breakdown of all components, different design decisions, such as
serial and unrolled implementations, and different ASIC and/or FPGA
implementation technologies can be found. Furthermore, some authors
report the results without interface, some with the interface, and in
some cases, \eg~\cite{firstlook}, CAESAR Hardware Applications
Programming Interface (API) for Authenticated Ciphers~\cite{caesarapi}
was used.  This paper describes how the algorithms for
{\ace}~\cite{acesub} and {\wage}~\cite{wagesub}, two LWC candidates,
were translated into hardware implementations; the parallel (unrolled)
implementations; and implementation results for four ASIC libraries.

Section~\ref{spec} describes the {\ace} and {\wage} permutations and
the sLiSCP sponge duplex mode~\cite{sliscp} that define the {\ace} and
{\wage} algorithms as specified in the {\ace} and {\wage} LWC
submissions~\cite{acesub,wagesub} and which represents work done by
the entire {\ace} and {\wage} teams.  Section~\ref{hwnopar} explores
how the permutation and mode were combined together into efficient
implementations in hardware and analyzes some of the key design
decisions in this process.  Section~\ref{hwpar} describes the parallel
implementations of {\ace} and \wage. Implementation technologies and
results are summarized in Section~\ref{results}.  The implementations
were synthesized using four different ASIC libraries, including 65\,nm,
90\,nm and 130\,nm technologies. {\ace} implementations range from a
throughput of 0.5 bits-per-clock cycle (bpc) and an area of 4210\,GE
(averaged across the four ASIC libraries) up to 4\,bpc and 7260\,GE.
{\wage} results range from 0.57\,bpc with 2920\,GE to 4.57\,bpc with
11080\,GE.

\section{Background}
\label{spec}

This section describes the {\ace} and {\wage} permutations and the
sLiSCP unified duplex sponge mode as specified in the NIST LWC
competition submissions~\cite{acesub,wagesub}. 
Both {\ace} and {\wage} permutations operate in a unified duplex
sponge mode~\cite{sliscp}. The 320\,bit {\ace} permutation offers both
AEAD and hashing functionalities, and the 259\,bit {\wage} permutation
supports AEAD functionality.  Because of the similarities between {\ace}
and {\wage}, this section begins with a short description of {\ace}
and {\wage} permutations, followed by a discussion on the unified
duplex sponge mode for both schemes, highlighting some differences.

\subsection{{\ace} Permutation}\label{spec-ace}

\begin{figure}[htbp]
  \centering
  \includegraphics[scale=0.6]{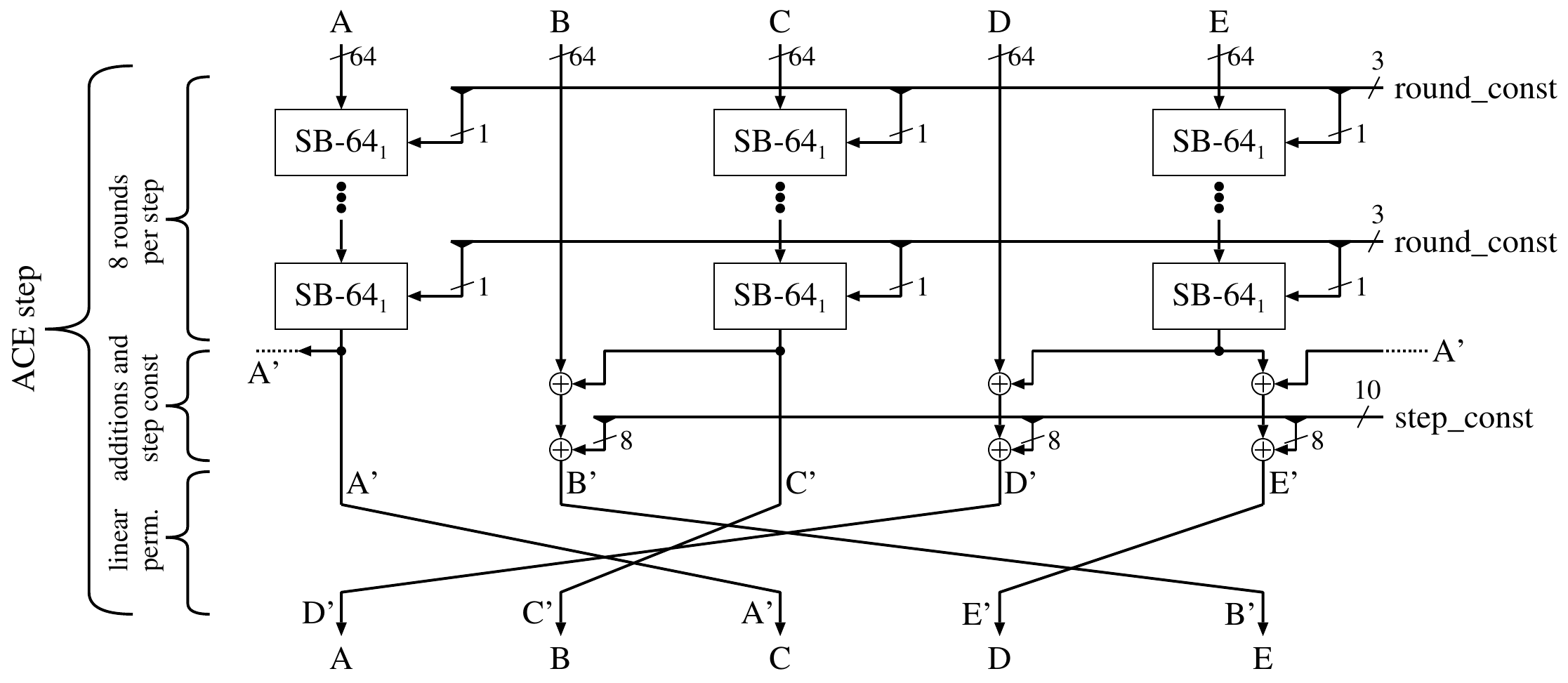}

  \caption{{\ace} step}
  \label{fig:ace-step}
  
\end{figure}

{\ace} has a 320 bit internal state, divided into five 64\,bit
registers, denoted \textsf{A}, \textsf{B}, \textsf{C}, \textsf{D}, and
\textsf{E}.  The {\ace} permutation consists of 16 steps
(Figure~\ref{fig:ace-step}), each of which consists of 8 rounds.
{\ace} uses the unkeyed reduced-round Simeck block
cipher~\cite{simeck_Yang2015} with a block size of 64, denoted Simeck
box {\simecki} for one round of the full 8-round S-box, as the
nonlinear operation. {\simeck} is a lightweight permutation,
consisting of left cyclic shifts, \textsc{and} and \textsc{xor} gates.
Each round is parameterized by a single bit of a round constant.

In each round, \simecki\ is applied to three registers \textsf{A},
\textsf{C} and \textsf{E}, with each using its own 1\,bit round
constant.  The round constants are generated by a 7\,bit LFSR run in a
3-way parallel configuration to produce one bit of each round constant
per clock cycle.

In each step, the outputs of \simecki\ are added to registers
\textsf{B}, \textsf{D} and \textsf{E}, which are further parameterized
by 8\,bit step constants .  The computation of step constants does not
need any extra circuitry, but rather uses the same LFSR as the round
constants: the three feedback values together with all 7 state bits
yield 10 consecutive sequence elements, which are then split into
three 8\,bit step constants. The step constants are used once every 8
clock cycles. The step function is then concluded by a permutation of
all five registers.  For the properties of \simeck, the choice of the
final permutation, and the number of rounds and steps refer
to~\cite{acesub}.

\subsection{{\wage} Permutation}
\label{spec-wage}

\wagename\ is a hardware oriented \aue\ scheme, built on top of the
initialization phase of the well-studied, {\lfsr} based, Welch-Gong
(\wg) stream cipher~\cite{wgestr, wggeneral}.  The {\wage} permutation
is iterative and has a round function derived from the \lfsr, the
decimated Welch-Gong permutation {\wgp}, and small S-boxes
{\sbx}. Details, such as differential uniformity and nonlinearity of
the {\wgp} and \sbx\ and selection of the {\lfsr} polynomial can be
found in~\cite{wagesub}.

\begin{figure}[htbp]
  \centering
  \includegraphics[scale=0.6]{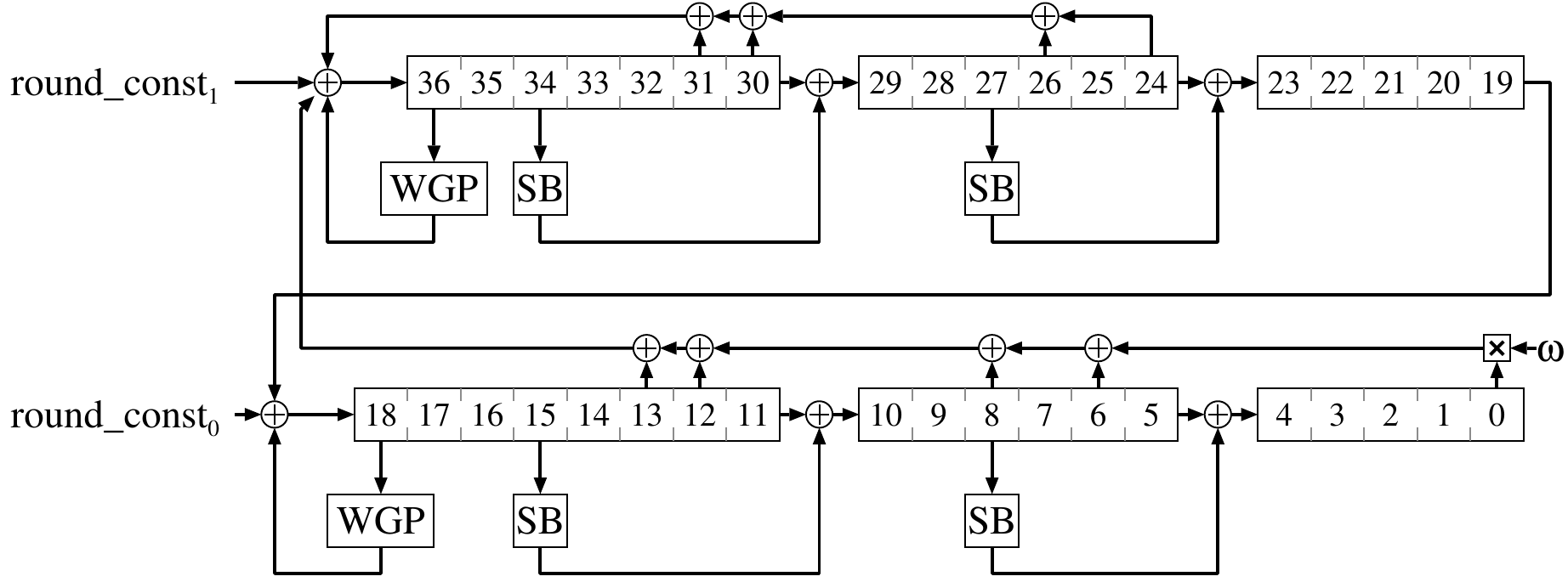}
  \caption{{\wage} round function}
  \label{fig:wage-round}
\end{figure}

Both {\lfsr} and {\wgp} are defined over \FF\, and the S-box is a
7\,bit permutation.  The {\lfsr} is defined by the feedback polynomial
\mbox{$y^{37}+y^{31}+y^{30}+y^{26}+y^{24}+y^{19}+y^{13}+y^{12}+y^8+y^6+\omega$},
which is primitive over \FF. The 37 stages of the {\lfsr} also
constitute the internal state of {\wage}.

The {\wage} permutation is iterative, and repeats its round function
(Figure~\ref{fig:wage-round}) 111 times. In each round, 6 stages of
the {\lfsr} are updated nonlinearly, while all the remaining stages
are just shifted. A pair of 7\,bit round constants is \textsc{xor}ed
with the pair of stages $(18, 36)$. Round constants are produced by a
7\,bit LFSR implemented in a 2-way parallel configuration.

\subsection{The Unified Duplex Sponge Mode }

\begin{figure}[htbp]
	\centering
        \subfigure[Authenticated encryption]{
  	  \includegraphics[scale=0.5]{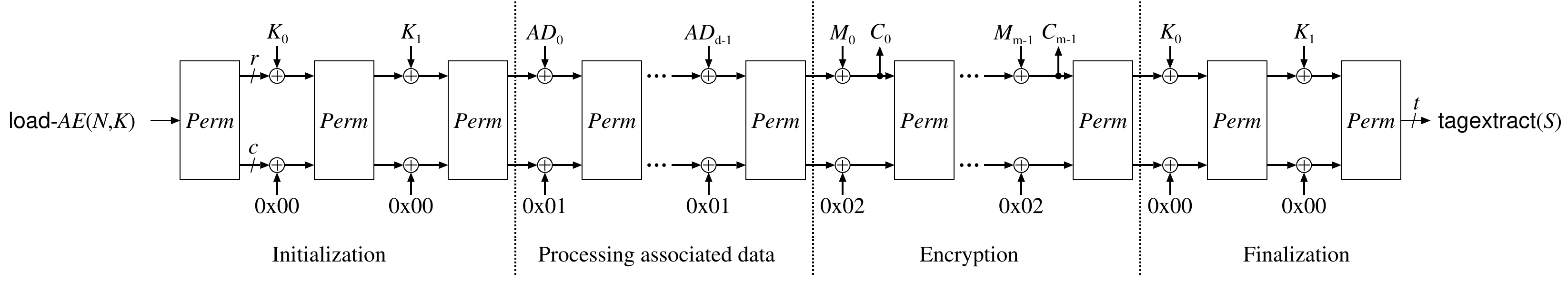}
        }
        
        \subfigure[Verified decryption]{
  	  \includegraphics[scale=0.5]{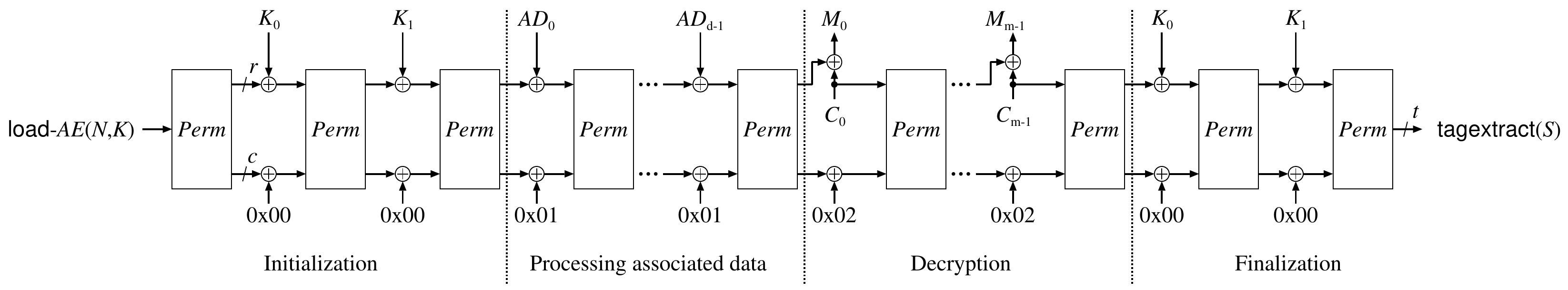}
        }

	\caption{Schematic diagram of the AEAD algorithm, where {\it Perm} is {\ace} or {\wage} permutation respectively}
	\label{fig.ae}
\end{figure}

\aceae\ and \wageae\ use  the unified duplex sponge mode  from sLiSCP
\cite{sliscp} (Figure~\ref{fig.ae}). The phases for encryption and decryption are:  initialization, processing of associated data,  encryption 
 (Figure~\ref{fig.ae}(a)) or decryption (Figure~\ref{fig.ae}(b)), and finalization.  Figure~\ref{fig.ae} also shows the domain separators for each phase.
The internal state is divided into a capacity part $S_c$ (256\,bits
for {\aceae} and 195\,bits for {\wage}) and a 64\,bit rate $S_r$, which for:  
\begin{itemize}
\item \aceae\ consists of bytes A[7], A[6], A[5], A[4], C[7], C[6], C[5], C[4] 
\item \wageae\ consists of the  0-th bit of stage $S_{36}$ (\ie, $S_{36,0}$)  and all bits of stages $S_{35}, S_{34}, 
S_{28}, S_{27}$,  $S_{18}, S_{16}, S_{15}, S_{9}$ and $S_{8}$ 
\end{itemize}

The input data (associated data $AD$, message $M$ or ciphertext $C$)
is absorbed (or replaced) into the rate part of the internal state.
If the input data length is not a multiple of 64, padding with
(10$^*$) is needed.  In Figure~\ref{fig.ae}, $d$ denotes the number of
64\,bit blocks of $AD$ and $m$ the number of 64\,bit blocks of $M$ and
$C$ after padding. Refer to~\cite{acesub, wagesub} for further padding
rules. No padding is needed during initialization and finalization
because the key, nonce, and IV are all 128\,bits, which is exactly 2
blocks. With the exception of tag extraction, both schemes generate an
output only during the encryption and decryption phases: the 64\,bit
output block is obtained by the \textsc{xor} of the current input and
rate.

The loading and tag-extraction functions are straightforward for \ace,
but more complicated for {\wage}.  For {\ace}, the key is loaded into
registers \textsf{A} and \textsf{C}, the nonce into \textsf{B} and
\textsf{E}, and the register \textsf{D} is loaded with zeros. The
{\ace} $\textsf{tagextract}(S)$ extracts the 128\,bit tag from
registers \textsf{A} and \textsf{C}.  For {\wage}, the loading and tag
extraction functions were designed to take advantage of the shifting
nature of the \lfsr, which will be discussed in more detail in
Section~\ref{hw-wage}.

\begin{figure}[htbp]
	\centering
	\includegraphics[scale=0.5]{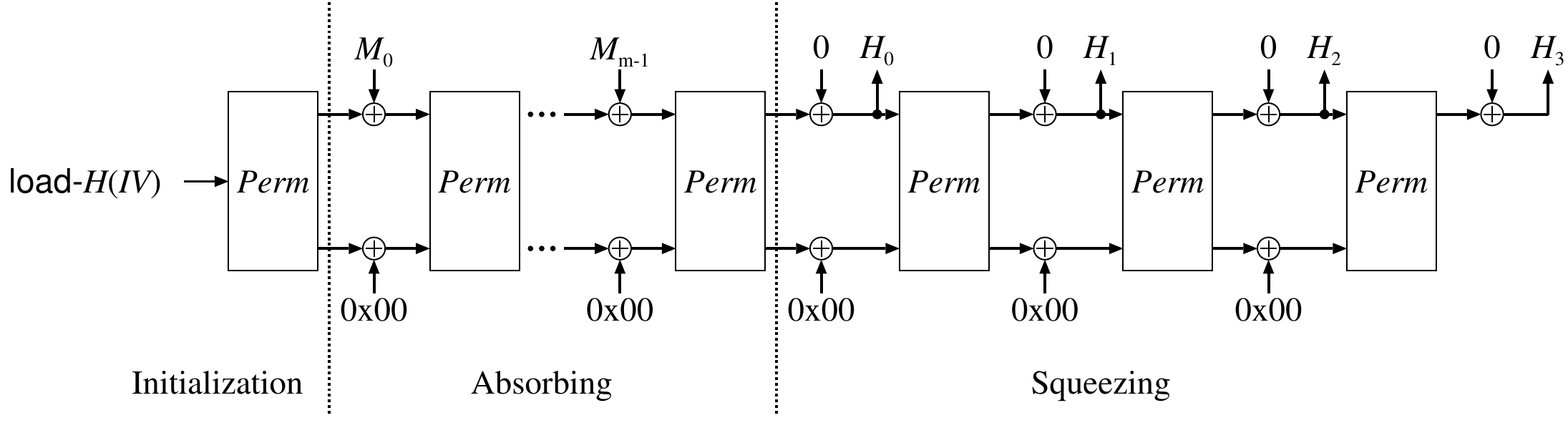}
	\caption{Schematic diagram of the  \acehash\ algorithm, where {\it Perm} is {\ace} permutation}
	\label{fig.hash}
\end{figure}

The {\ace} HASH functionality is shown in Figure~\ref{fig.hash}. The
only input is the message $M$. Since the hash has a fixed length
of 256\,bits, the length of the squeezing phase is fixed.  \acehash\
is unkeyed, and the state is loaded with a fixed initialization vector
IV. More specifically, the function $\textsf{load-}\mathcal{H}(IV)$
loads the state bytes B[7], B[6] and B[5] with bytes 0x80, 0x40, and
0x40 respectively, and sets all other state bits to zero.

\section{Hardware Implementations}
\label{hwnopar}

This section describes and analyses the hardware implementations of
{\ace} and {\wage}.  The next section describes the parallel
implementations, followed by results in Section~\ref{results}.
Section~\ref{hw-principles} summarizes the principles underlying the
hardware design and the interfaces for the ciphers.
Sections~\ref{hw-ace} and \ref{hw-wage} describe {\ace} and {\wage}.
Section~\ref{sec:decisions} summarizes some of the key
hardware-oriented design decisions.

\subsection{Hardware Design Principles and Interface with the
  Environment}
\label{hw-principles}

\begin{figure}[htbp]
  \begin{minipage}[t]{0.4\linewidth}
    \centering
    \includegraphics[scale=0.65]{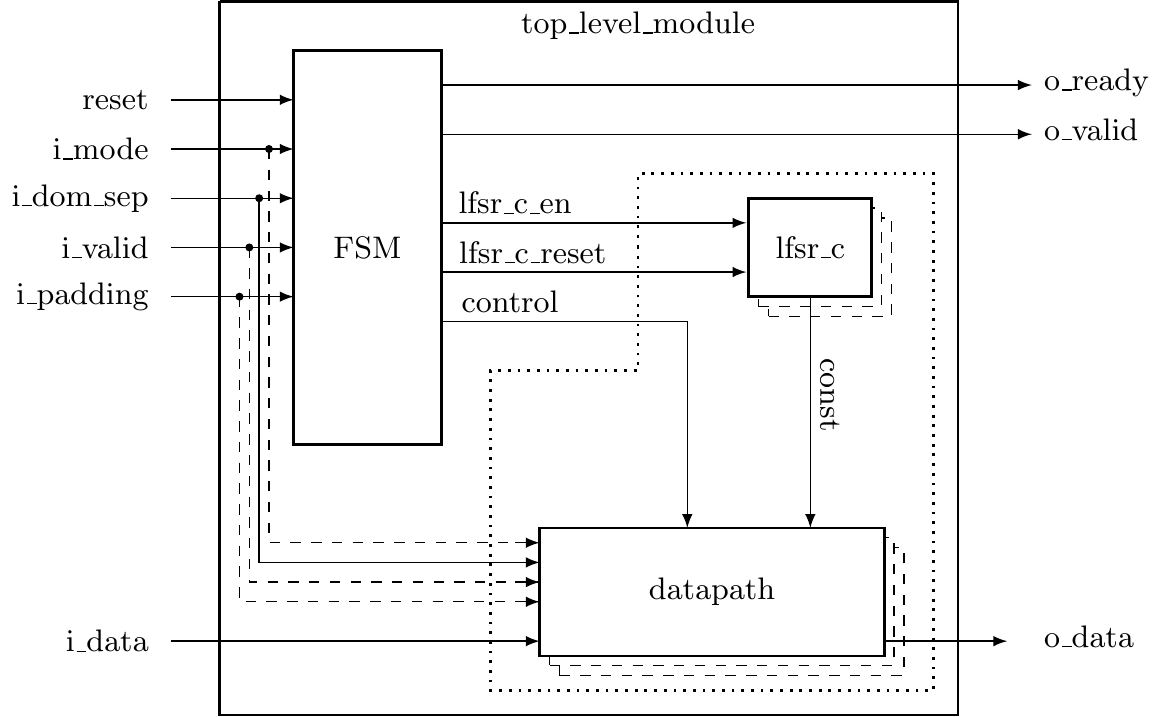}
    \caption{Top-level module and interface}
    \label{fig.arch}
  \end{minipage}
  \hfill
  \begin{minipage}[t]{0.55\linewidth}
    \centering
    \includegraphics[scale=0.5]{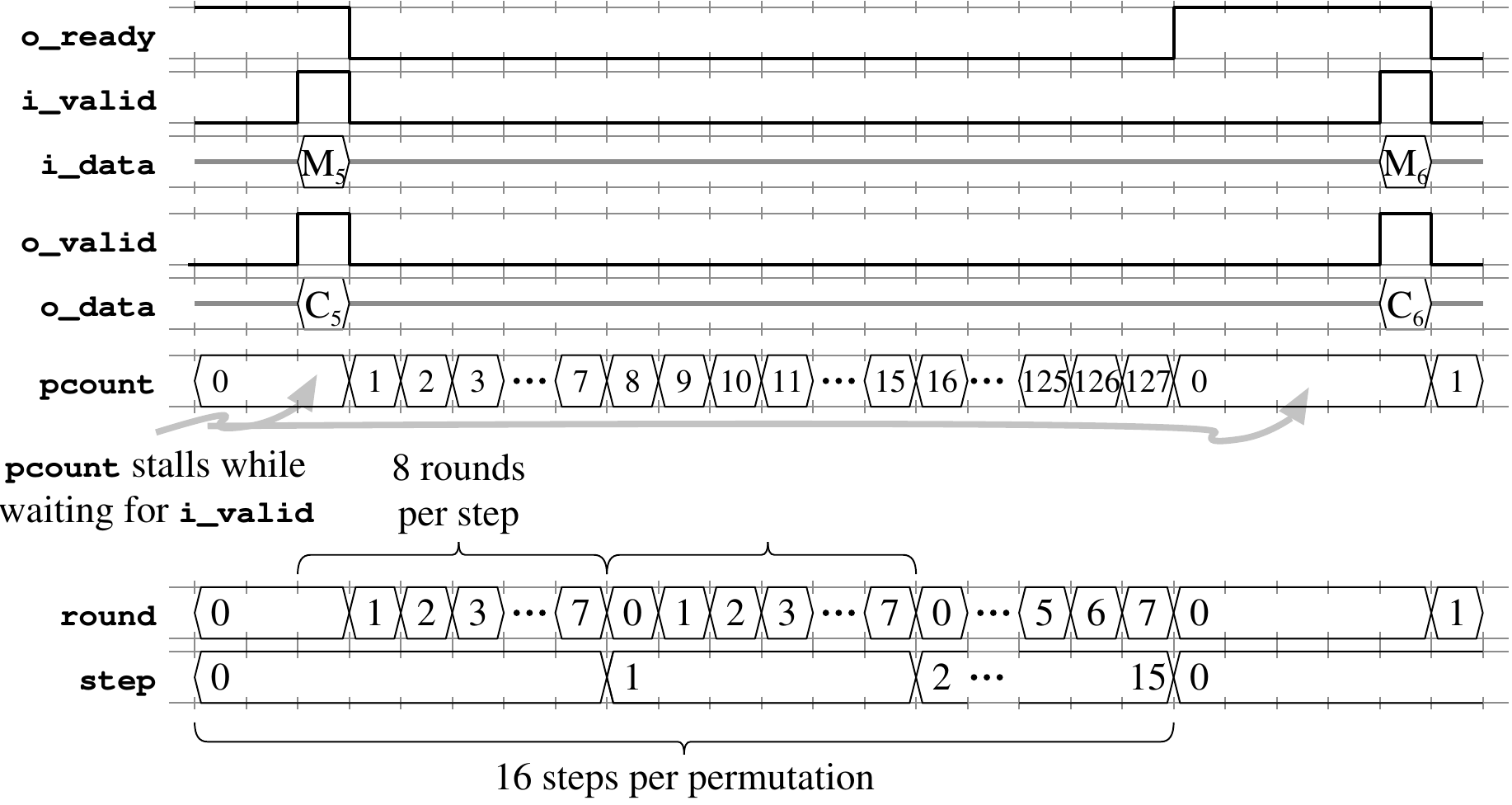}
    \caption{Timing diagram for {\ace} during encryption}
    \label{fig.timing}
  \end{minipage}
\end{figure}

\begin{enumerate}\itemsep-2pt  \parsep-4mm \vspace{-0.5em}
 
\item {\bf Multi-functionality module.}  The system should include all
  supported operations in a single module  (Figure~\ref{fig.arch}),
   because lightweight applications cannot afford the extra area for separate
  modules.

\item {\bf Single input/output ports.}  In small devices, ports can be
  expensive.  To ensure that {\ace} and {\wage} are not biased in favour of
  the system, at the expense of the environment, the ciphers have one
  input and one output port (Table~\ref{tbl.interface}).  That being said,
  the authors agree with the proposed lightweight cryptography
  hardware API's~\cite{lwc-hw-api} use of separate public and private
  data ports and will update implementations accordingly.

\item {\bf Valid-bit protocol and stalling capability.}  The
  environment may take an arbitrarily long time to produce any piece
  of data. For example, a small microprocessor could require multiple
  clock cycles to read data from memory and write it to the system's
  input port.  The receiving entity must capture the data in a single
  clock cycle (Figure~\ref{fig.timing}). In reality, the environment can stall as well.
  In the future, {\ace} and
  {\wage} implementations will be updated to match the proposed
  lightweight cryptographic hardware API's use of a valid/ready protocol for
  both input and output ports.
	
\item {\bf Use a ``pure register-transfer-level'' implementation
    style. } In particular, use only registers, not latches;
  multiplexers, not tri-state buffers; synchronous, not asynchronous
  reset; no scan-cell flip-flops; clock-gating is used for power and
  area optimization.
        
\end{enumerate}
\vspace{-0.5em}

Since both {\ace} and {\wage} use a unified sponge duplex mode, they share a common interface with the environment (Table~\ref{tbl.interface}). 
The environment separates the associated data and the message/ciphertext, and performs  padding if necessary. The domain separators shown in Figure~\ref{fig.ae} are provided by the environment and serve as an indication of the phase change for AEAD functionality. For {\acehash}, the phase change is indicated by the change of the
\textsf{i\_mode(0)} signal, as shown in Table~\ref{tab.acehwmodes}. The hardware is unaware of the lengths of individual phases, hence no internal counters for the number of processed blocks are needed.   

\begin{table}[htbp]
  \begin{minipage}[t]{0.48\linewidth}
    \footnotesize
    \centering

    \caption{Interface signals}
    \label{tbl.interface}
    
    \begin{tabular}[t]{ |l l|}  \hline 
      \textbf{Input signal} & \textbf{Meaning} \\\hline
      \texttt{reset} & resets the state machine \\
      \texttt{i\_mode} & mode of operation \\
      \texttt{i\_dom\_sep} & domain separator \\
      \texttt{i\_padding} & the last block is padded \\
      \texttt{i\_data} &  input data \\
      \texttt{i\_valid} &  valid data on \texttt{i\_data} \\ \hline 
    \hline 

      \textbf{Output signal} & \textbf{Meaning}\\\hline
        \texttt{o\_ready} & hardware is ready \\
        \texttt{o\_data} & output data\\
      \texttt{o\_valid} & valid data on \texttt{o\_data} \\ \hline 
    \end{tabular}

  \end{minipage}
  \hfill
  \begin{minipage}[t]{0.48\linewidth}
    \footnotesize
    \centering

    \captionof{table}{Modes of operation}
    \label{tab.acehwmodes}
    \vspace{-1em}

     \begin{tabular}{|c c|  c |c|} \hline 
      \multicolumn{2}{|c|}{\texttt{i\_mode}} & & \\
      \cline{1-2}
       (1) &(0) & Mode & Operation or phase \\ \hline
        0 & 0 & {\aceenc} & Encryption \\
        0 & 1 & {\acedec} & Decryption \\
        1 & 0 & {\acehash}  & Absorb     \\
        1 & 1 & {\acehash}  & Squeeze    \\
        \hline  \hline

  -&0	&\wageenc &Encryption \\
  -&1    &\wagedec &Decryption \\
  \hline

    \end{tabular} 
  \end{minipage}
\end{table}

The top-level module, shown in Figure~\ref{fig.arch},  is also very similar for both {\ace} and \wage. 
It depicts the interface signals from Table~\ref{tbl.interface}, with only slight differences in bitwidths. Figure~\ref{fig.timing} shows the timing diagram for {\ace} during the encryption phase of message blocks $M_5$ and $M_6$, which clearly shows the valid-bit protocol. The first five lines show the top-level interface signals and line six shows the value of the permutation counter {\tt pcount}, which is a part of the {\ace} finite state machine (FSM) and keeps track of the $128$ clock cycles needed for one {\ace} permutation. 
After completing the previous permutation, the top-level module
asserts {\tt o\_ready} to signal to the environment that an {\ace}
permutation just finished and new data can be accepted. The
environment replies with a new message block $M_5$ accompanied by an
{\tt i\_valid} signal. The hardware immediately encrypts, returns
$C_5$ and asserts {\tt o\_valid}. This clock cycle is also the first
round of a new {\ace} permutation and the {\tt o\_ready} is
deasserted, indicating that the hardware is
busy. Figure~\ref{fig.timing} shows the {\ace} hardware remaining busy
({\tt o\_ready = 0}) for the duration of one {\ace} permutation. When
{\tt pcount} wraps around from 127 to 0, the hardware is again idle
and ready to receive new input, in this case $M_6$. A few more details
about the use of {\tt pcount} will follow in
Subsection~\ref{hw-ace}. The interaction between the top-level module
and the environment during the encryption phase of {\wage} is very
similar, with 111 clock cycles for the completion of one
permutation. More significant differences for the interaction with the
environment arise during loading, tag extract and of course \acehash.

\subsection{ACE Datapath}
\label{hw-ace}

\begin{figure}[htbp]
  \begin{minipage}{0.69\linewidth}
    \subfigure[The {\ace} datapath]{
      \includegraphics[scale=0.55]{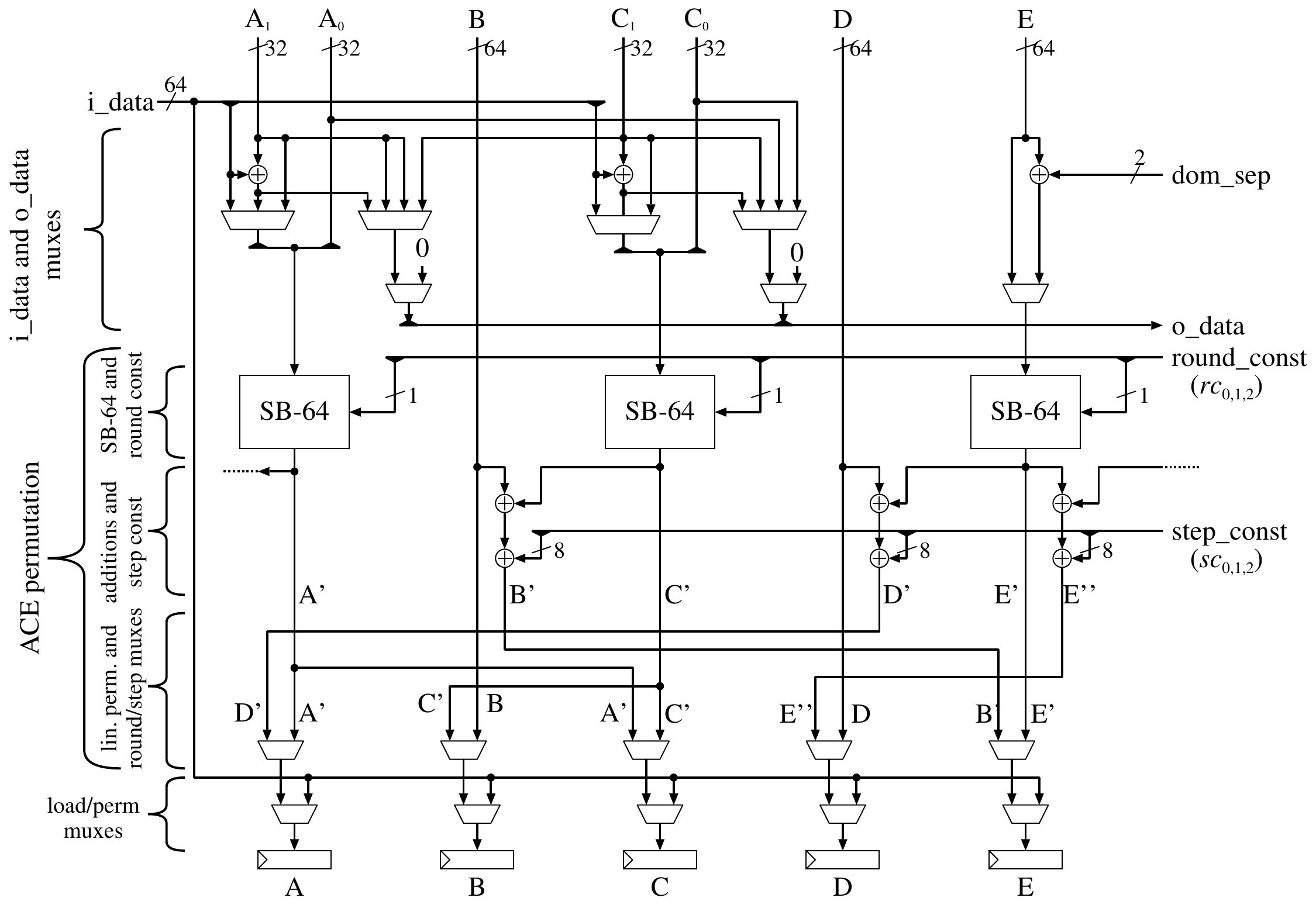}
    }
  \end{minipage}
  \begin{minipage}{0.29\linewidth}
    \subfigure[Parallelization $p=4$ segment for registers A and B]{
      \includegraphics[scale=0.55,origin=t]{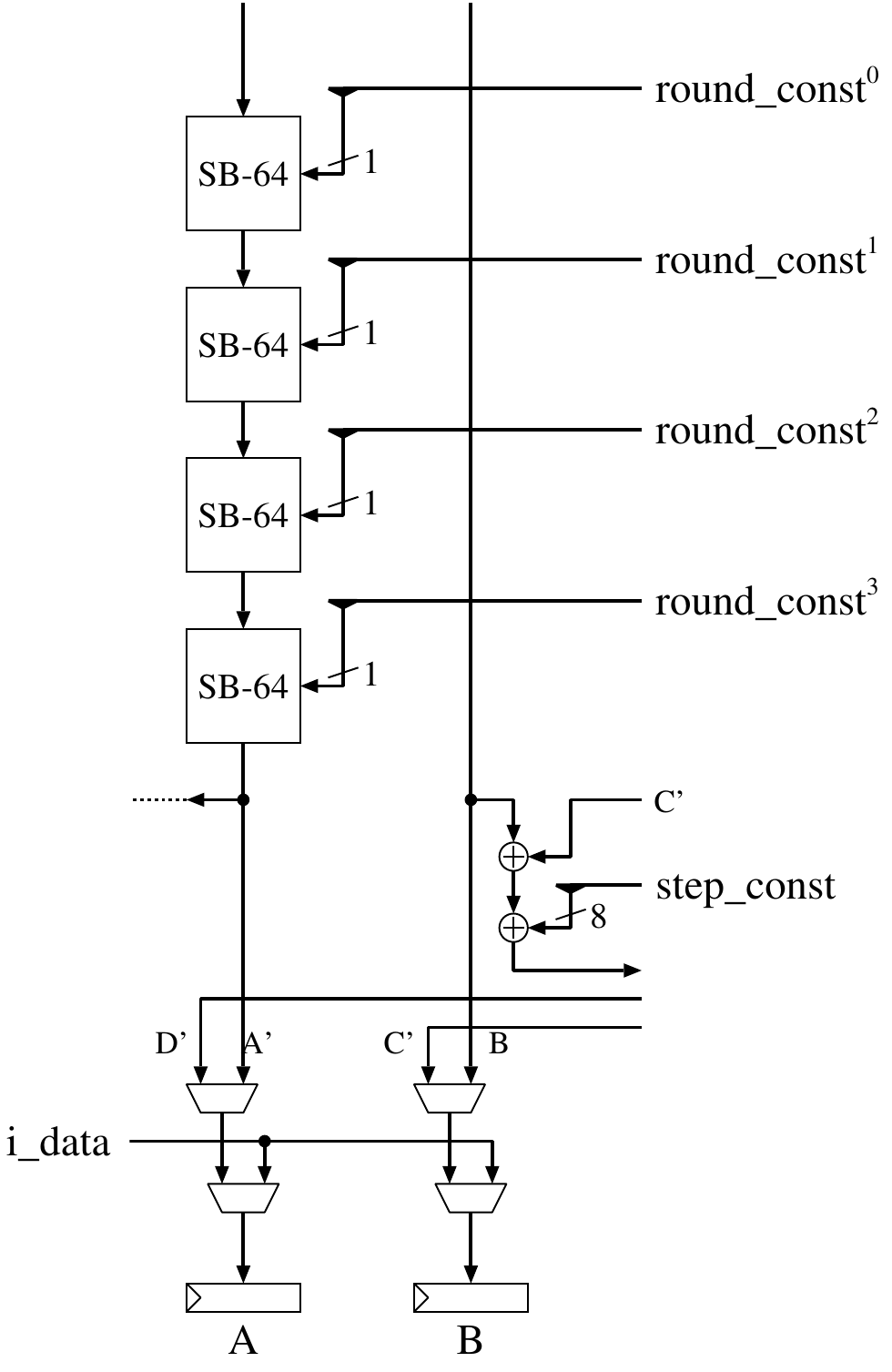}
    }
  \end{minipage}
  
  \captionof{figure}{The {\ace} datapath and parallelization}
  \label{fig.acedp}
\end{figure}

Figure~\ref{fig.acedp}(a) shows the {\ace} datapath. The top and
bottom of the figure depict the five 64\,bit registers \textsf{A},
\textsf{B}, \textsf{C}, \textsf{D} and \textsf{E}.

The support for the different phases in the mode (\eg, load,
initialization, encryption, \etc) imposes a variety of multiplexers
onto the core datapath for performing the round computation.  The
multiplexers are controlled by the mode and the counter
\texttt{pcount}.  Near the top of Figure~\ref{fig.acedp}(a), are
multiplexers to determine the input to the round function and the data
that is sent to the output.  Next, the output multiplexers are needed
to accommodate encryption/decryption and tag generation for \aceae\
and squeezing for \acehash. Furthermore, the output is forced to 0
during normal operation.  The middle of the figure includes {\simecki}
for the round function.  Below {\simecki} is the hardware for the step
function: \textsc{xor} gates for the addition and step constant and
the linear permutation.  The first row of multiplexers at the bottom
is to select between doing a round vs.\@ step; the bottom row is to
select between loading (key, nonce, IV) or doing a normal (round,
step, permutation) computation.

\subsection{WAGE Datapath}
\label{hw-wage}

Because of {\it the shifting nature of the \lfsr}, which in turn
affects loading, absorbing and squeezing, the {\wage} datapath is
slightly more complicated than the {\ace} datapath and hence is
explained in two levels:

\begin{figure*}[htbp]
	\centering
	\includegraphics[scale=0.4]{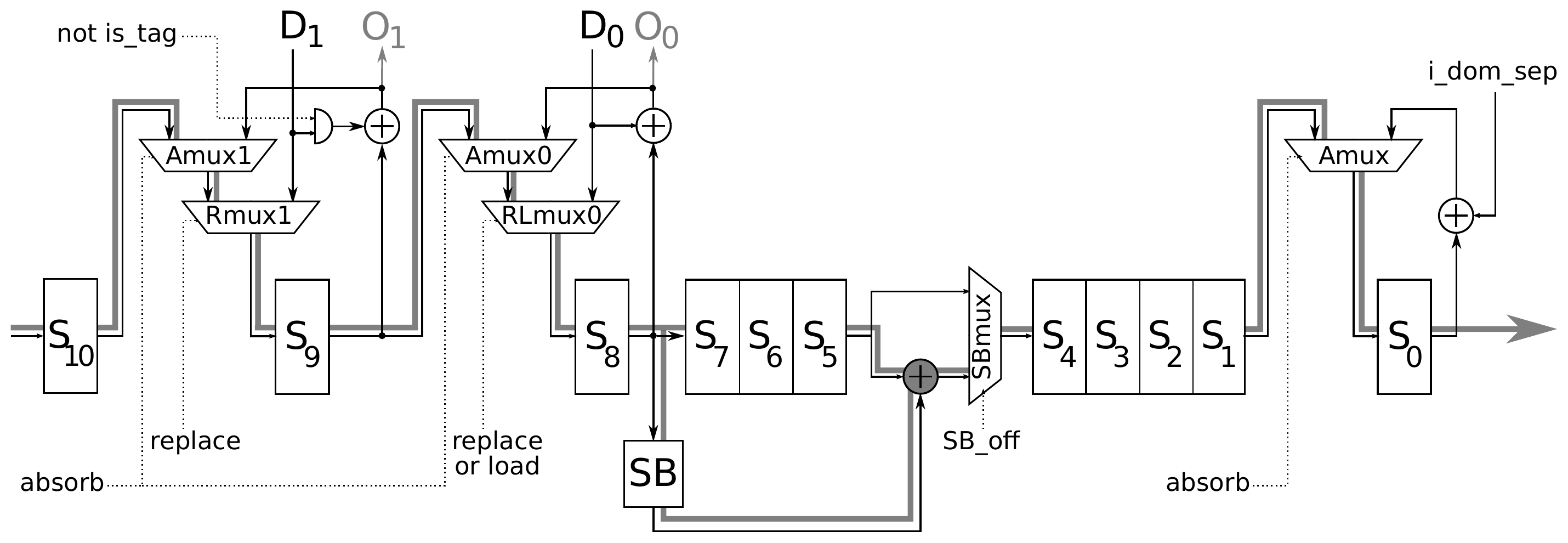}
	\caption{The {\wagelfsr} stages $S_{0},\dots, S_{10}$ with multiplexers, \textsc{xor} and \textsc{and} gates for the sponge mode}
	\label{fig.wagelfsr}
\end{figure*}

\begin{enumerate}\itemsep-2pt  \parsep-4mm \vspace{-0.5em}
  \item Implementation of the permutation
    \begin{itemize}\itemsep-2pt  \parsep-4mm \vspace{-0.5em}

      \item \wgphw\ module implementing \wgp: For smaller fields like
        \FF, the \wgphw\ area, when implemented as a constant array in
        VHDL/Verilog (\ie, as a look-up table) is smaller than when
        implemented using components such as multiplication and
        exponentiation to powers of two~\cite{wg5,wg7}. However, the
        {\wgp} is not implemented in hardware as a memory array, but rather
        as a net of \textsc{and}, \textsc{or}, \textsc{xor} and \textsc{not} gates,
        derived and optimized by the synthesis tools.

      \item \sbx\ module: The function is defined as 5 iterations of
        smaller functions, but is implemented in unrolled fashion,
        (\ie, as purely combinational logic).

      \item \lfsrc: The {\lfsrc} for generating the round constants was
        implemented in a 2-way parallel fashion. It has only 7 1\,bit
        stages and two \textsc{xor} gates for the two feedback
        computations.

    \end{itemize}

  \item Support for the sponge mode.  Figure~\ref{fig.wagelfsr} shows details for stages
    $S_{0}, \dots, S_{10}$. The grey line represents the path for
    normal operation during the {\wage} permutation.  The additional
    hardware for the entire {\wagelfsr} is listed below, with examples
    in brackets referring to Figure~\ref{fig.wagelfsr}.

    \begin{itemize}\itemsep-2pt  \parsep-4mm \vspace{-0.5em}
      \item The 64\,bit {\tt i\_data} is padded with zeros to 70 bits,
      then fragmented into 7\,bit {\wagelfsr} inputs $D_k$,
      $k=0,\dots,9$, corresponding to the rate stages $S_r$. For each
      data input $D_k$ there is a corresponding 7\,bit data output
      $O_k$. ( $D_1,O_1$ and $D_0,O_0$ in Figure~\ref{fig.wagelfsr}).
 
    \item 10 \textsc{xor} gates must be added to the $S_r$ stages to
      accommodate absorbing, encryption and decryption (\textsc{xor}s
      at stages $S_9$,$S_8$).

    \item 10 multiplexers to switch between absorbing and normal
      operation ({\tt Amux1}, {\tt Amux0} at $S_9$,$S_8$).

    \item An
      \textsc{xor} and a multiplexer are needed to add the domain
      separator {\tt i\_dom\_sep} ({\tt Amux} at $S_0$).

    \item To
      replace the contents of the $S_r$ stages, 10 multiplexers are
      added ({\tt Rmux1} at stage $S_9$)

    \item Instead of additional
      multiplexers for loading, the existing {\tt Rmux$k$},
      $k=9,5,4,3,0$, multiplexers are now controlled by {\tt replace
        or load} and labelled {\tt RLmux$k$}, (see {\tt RLmux0} on
      $S_8$). Since all non-input stages must keep their previous
      values, an enable signal {\tt lfsr\_en} is needed.

    \item Three
      7\,bit \textsc{and} gates to turn off the inputs $D_6, D_3$ and
      $D_1$ (\textsc{and} at $D_1$).

    \item Four multiplexers are needed
      to turn off the \sbxhw\ during loading and tag extraction ({\tt
        SBmux} at $S_4$).

    \item The total hardware cost to support
      the sponge mode is: 24 7\,bit and one 2\,bit multiplexers, 10 7\,bit
      and one 2\,bit \textsc{xor} gates, three 7\,bit \textsc{and}
      gates.

    \end{itemize}

\end{enumerate}

As mentioned in Section 2, special care was given to the design of
loading and tag-extract.  The existing data inputs $D_k$ are reused
for loading, and the outputs $O_k$ for tag extraction.  The
{\wagelfsr} is divided into five loading regions using the inputs
$D_k$, $k=9,5,4,3,0$. For example, the region $S_{0},\dots,S_{8}$ in
Figure~\ref{fig.wagelfsr} is loaded through input $D_0$, however,
instead of storing $D_0\oplus S_8$, the $D_0$ data is fed directly
into $S_{8}$, \ie, the {\tt RLmux0} disconnects the {\tt Amux0}
output.  The remaining stages in this region are loaded by shifting,
which requires the {\tt SBmux} at $S_4$. Note that there is no need to
disconnect the two {\wgp}, because they are automatically disabled by
loading through $D_9$ and $D_4$, located at stages $S_{36}$ and
$S_{18}$ respectively.   The tag is
extracted in a similar fashion as loading, but from the data output
$O_k$ at the end of a particular loading region, \eg, the region
$S_{9},\dots,S_{16}$, loaded through $D_3$, is extracted through
$O_1$. The longest tag extraction region is of length 9, which is the
same as the longest loading region.

\subsection{Hardware-Oriented Design Decisions and Analysis}
\label{sec:decisions}

The design process for {\ace} and {\wage} tightly integrated cryptanalysis and
hardware optimizations.  A few key hardware-oriented decisions
are highlighted here;  more can be found in the design rationale
chapters of~\cite{acesub,wagesub}.

Functionally, it is equivalent for the boundary between phases to
occur either before or after the permutation. For {\ace} and {\wage},
 the boundary was placed \emph{after} the permutation updates the 
 state register.  This means
that the two-bit domain separator is sufficient to determine the value
of many of the multiplexer select lines and other control signals.
All phases that have a domain separator of \texttt{"00"} have the same
multiplexer select values.  The same also holds true for
\texttt{"01"}.  Unfortunately, this cannot be achieved for
\texttt{"10"}, because encryption and decryption require different
control signal values, but the same domain
separator. Using the domain separator to signal the transition 
between phases for encryption and decryption also simplifies the control circuit.
For hashing, the change in phase is indicated by the 
{\tt i\_mode} signal.

In applications where the delay through combinational circuitry is not
a concern, such as with lightweight cryptography, where clock speed is
limited by power consumption, not by the delay through combinational
circuitry, it is beneficial to lump as much combinational
circuitry as possible together into a single clock cycle.  This
provides more optimization opportunities for the synthesis tools than
if the circuitry was separated by registers.  For this reason, 
the {\ace} datapath was designed so that the input and output multiplexers, one round
of the permutation, and state loading multiplexers
together form a purely combinational circuit, followed by the
state register. 

\begin{figure}[htbp]
    \includegraphics[scale=0.6]{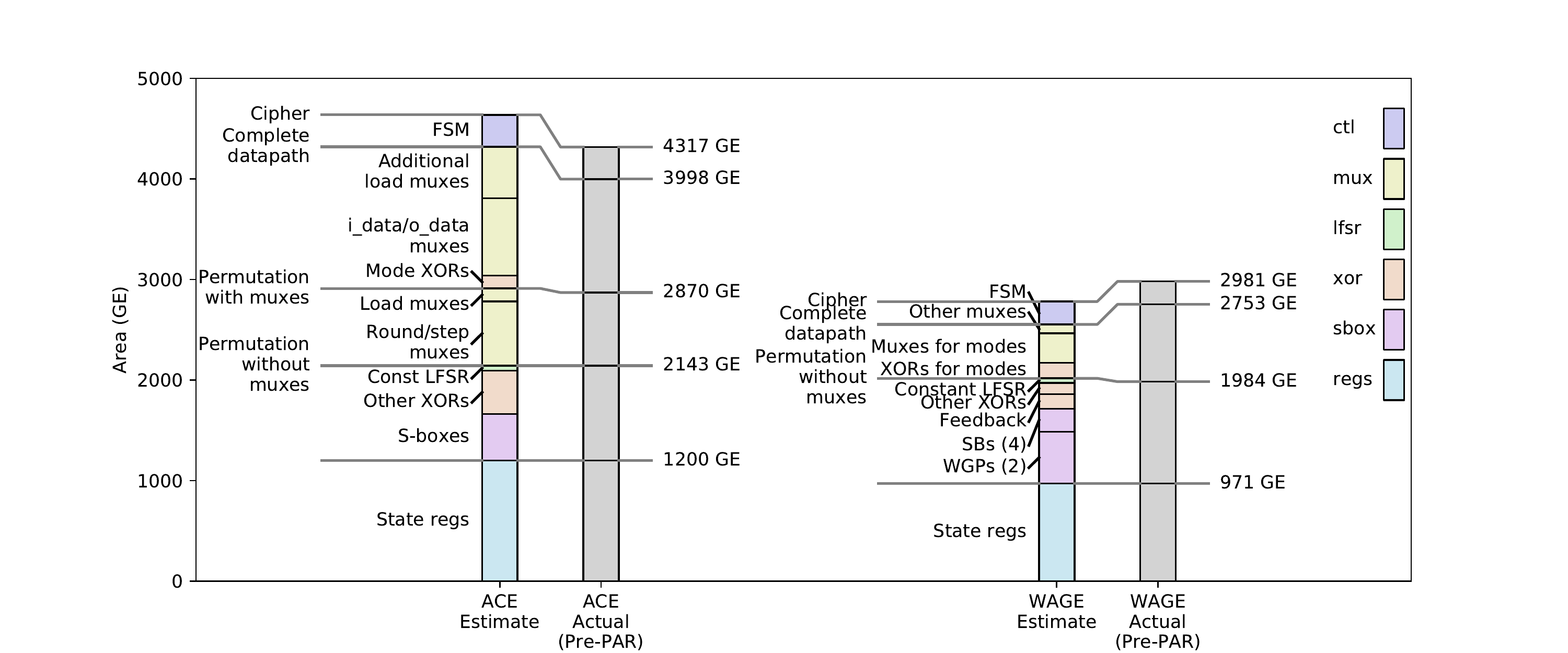}
    \caption{Area analysis}
    \label{fig:area-analysis}
\end{figure}

Figure~\ref{fig:area-analysis} shows the area analysis and comparison
for {\ace} and {\wage}.  The coloured columns show the estimated area
for different components based on counting gates from the mathematical
definitions.  The grey columns show the actual area from logical
synthesis using a compilation script with minimal optimizations so as
to preserve structural hierarchy.  Table~\ref{hwresults} shows the
final results with optimized synthesis scripts and after physical
synthesis (post place-and-route).  Figure~\ref{fig:area-analysis}
shows that {\ace} and {\wage} have relatively similar area for the
permutation with multiplexers: 2143\,GE for {\ace} and 1984\,GE for
{\wage}, making {\ace} 8\% larger.  The larger area of {\ace} is due
mostly to the large number of multiplexers used to support the mode
and input/output.  

The synthesized area of {\ace} is less than the estimated area, while
the opposite is true for {\wage}.  As can be seen in comparing
Figures~\ref{fig.acedp} and \ref{fig.wagelfsr}, the {\wage} datapath
is distributed into a large number of small lumps separated by
registers, while the {\ace} datapath is relatively monolithic, which
gives the synthesis tools more opportunities for optimizations.

\section{Parallel Implementations}
\label{hwpar}

\subsection{Parallelization in General}

Both ciphers can be parallelized (unrolled) to execute multiple rounds per clock
cycle, at the cost of increased area. In the top-level schematic in
Figure~\ref{fig.arch}, the dashed stacked boxes indicate
parallelization. The FSM is parameterized with parameter $p$ and used
for un-parallelized ($p{=}1$) and parallelized ($p{>}1$)
implementations. Other components are replicated to show $p$ copies,
with $p{=}3$ in Figure~\ref{fig.arch}. Such a representation is
symbolic;  parallelization is applied only to the permutation,
not the entire datapath. The interface with the environment remains
the same.

\subsection{ACE}
\label{sec:ace-par}

The $p{=}1$ un-parallelized {\ace} permutation performs a single round
per clock cycle, which implies 8 clock cycles per step. Parallel (\ie,
unrolled) versions perform $p$ rounds per clock cycle, and were
implemented for divisors of 8 (\ie\ $p=2,4,8$). The {\ace} permutation
could be parallelized further, \eg\ two or more steps in a single
clock cycle.  Figure~\ref{fig.acedp}(b) shows the example $p{=}4$ for
registers \textsf{A} and \textsf{B}, with $p{=}4$ copies of \simecki\
connected in series. Each \simecki\ has its own round constant
$\rc_0^k$, $k=0,\dots,p-1$.  The round vs.\@ step multiplexers are
still needed, and can be removed only for values of $p$, that are
multiples of 8.  Also note the step constant indicated as
$\sc_0^{p-1}$.  For $p{=}4$ a step is concluded in 2 clock
cycles. However, this requires a modification to the \lfsrc, which
must now generate $p\cdot3$ round constant bits $\rc_j^k$, $j=0,1,2$,
$k=0,\dots,p-1$ per clock cycle.  The last cycle within a step
requires 7 additional bits, which together with $\rc_j^{p-1}$ yield 10
bits for the step constant generation $\sc_j^{p-1}$. In the case
$p{=}4$ the {\lfsrc} must generate 12 constant bits in the first cycle
and 19 constant bits in the second clock cycle of the step, which are
then used for $\rc_j^k$ and $\sc_j^k$. For the extra constant bits,
the {\lfsrc} feedback was replicated (\ie, $(p-1)\cdot 3$ feedbacks in
addition to the original 3).

\subsection{WAGE}
\label{sec:wage-par}

\begin{figure*}[htbp]
	\centering
	\includegraphics[scale=0.48]{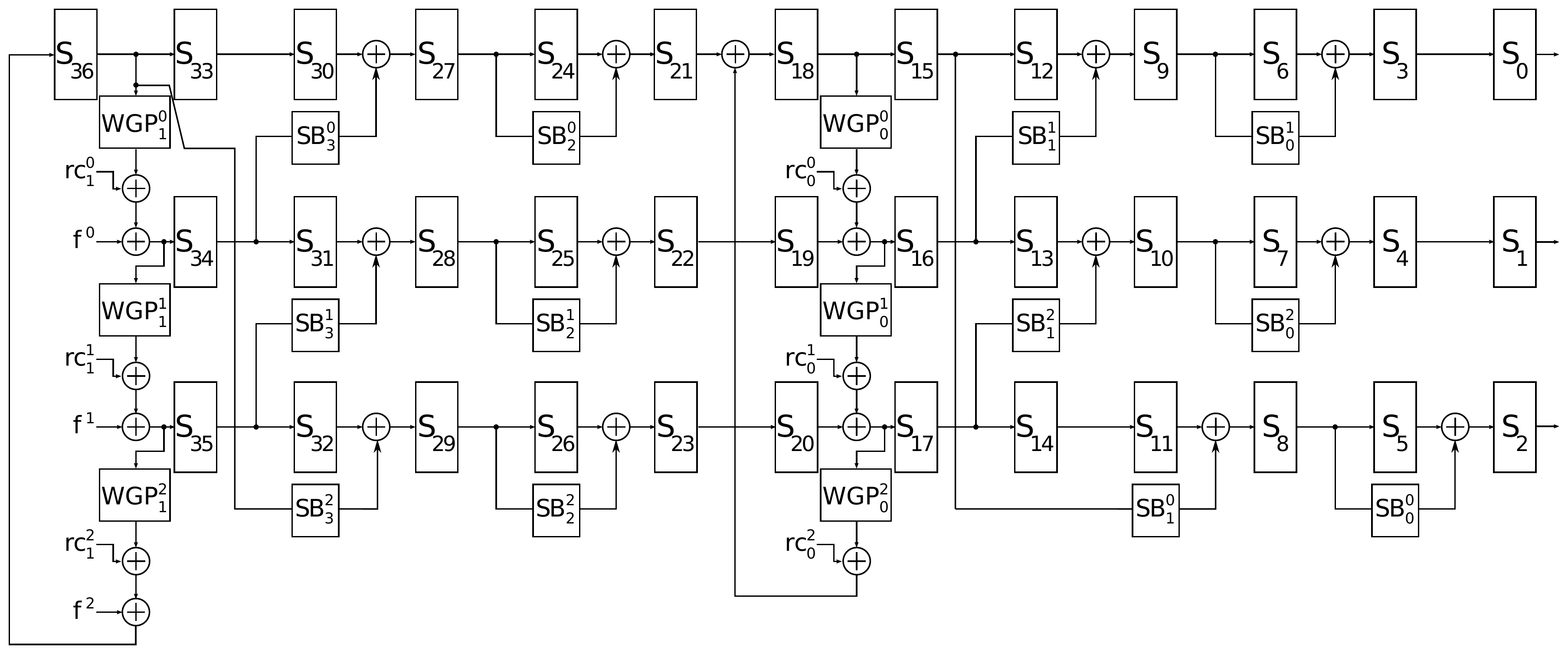}
	\caption{The {\wage} permutation with $p{=}3$}
	\label{fig.wagelfsr3}
\end{figure*}

{\wage} performs one clock cycle for the interaction with the
environment
(\ie\ absorbing or replacing the input data into the state) followed by 111 clock cycles of the {\wage}
 permutation.
Because 111 is divisible only by 3 and 37, the opportunities to parallelize
{\wage} appear rather limited.  However, by treating the absorption
or replacement of the input data into the internal state as an
additional clock cycle in the permutation, we increase the the length
of the permutation to 112 clock cycles.  Because 112 has many
divisors, this allows parallelism of $p=2,3,4,6,8$.  The cost is 
some additional multiplexers, because the clock cycle that loads data
has different behaviour than the normal clock cycles

Figure~\ref{fig.wagelfsr3} shows the 3-way parallel {\wagelfsr}
including all nonlinear components and their copies.  Multiplexers are
not replicated, and hence, are not shown.  For the components
$f, \rc, \wgp$ and $\sbx$ in Figure~\ref{fig.wagelfsr3}, the
superscript $k$ indicates the original ($k=0$) and the two copies
($k=1,2$).  Computation of the three feedbacks $f^k$ is not shown but
is conducted as
$f^k = S_{31+k}\oplus S_{30+k} \oplus S_{26+k} \oplus S_{24+k} \oplus
S_{19+k}\oplus S_{13+k}\oplus S_{12+k}\oplus S_{8+k}\oplus S_{6+k}
\oplus (\omega\otimes S_{0+k})$. Similar to \ace, the generation of
{\wage} round constants $\rc_{1}^k, \rc_{0}^k$ must be parallelized as
well.  For readability, the two {\wgp} were labelled
$\wgp_1^k, \wgp_0^k$, with $\wgp_1^0, \wgp_0^0$ being the original
\wgp s positioned at $S_{36}, S_{18}$, just like $\rc_1^0,
\rc_0^0$. Similarly, the \sbx s were also labelled
$\sbx_j^k$, $j=3,2,1,0$, in the decreasing order (\ie, $\sbx_3^0$ is the original \sbx\ with input $S_{34}$).

\section{Implementation Technologies and ASIC Implementation Results}
\label{results}

Logic synthesis was performed with Synopsys Design
  Compiler version P-2019.03 using the \texttt{compile\_ultra} command
  and clock gating.  Physical synthesis (place and route) and power
  analysis were done with Cadence Encounter v14.13 using a density of
  95\%. simulations were done in Mentor Graphics ModelSim SE v10.5c.
  The ASIC cell libraries used were ST Microelectronics 65\,nm
  CORE65LPLVT 1.25V, TSMC 65 nm tpfn65gpgv2od3 200c and tcbn65gplus
  200a at 1.0V, ST Microelectronics 90 nm CORE90GPLVT and CORX90GPLVT
  at 1.0V, and IBM 130nm CMRF8SF LPVT with SAGE-X v2.0 standard cells
  at 1.2V. Some past works have used scan-cell flip-flops to reduce
  area, because these cells include a 2:1 multiplexer in the flip-flop
  which incurs less area than using a separate multiplexer. 
 Scan-cell flip-flops were not used because their use as part of the
  design would prevent their insertion for fault-detection and hence,
  prevent the circuit from being tested for manufacturing faults. 
  Furthermore, chip enable signals were removed from all datapath registers, which are controlled by clock gating instead. This allows a further reduction of the implementation area. 

\begin{figure*}[htbp]
	\centering
	\includegraphics[scale=0.7]{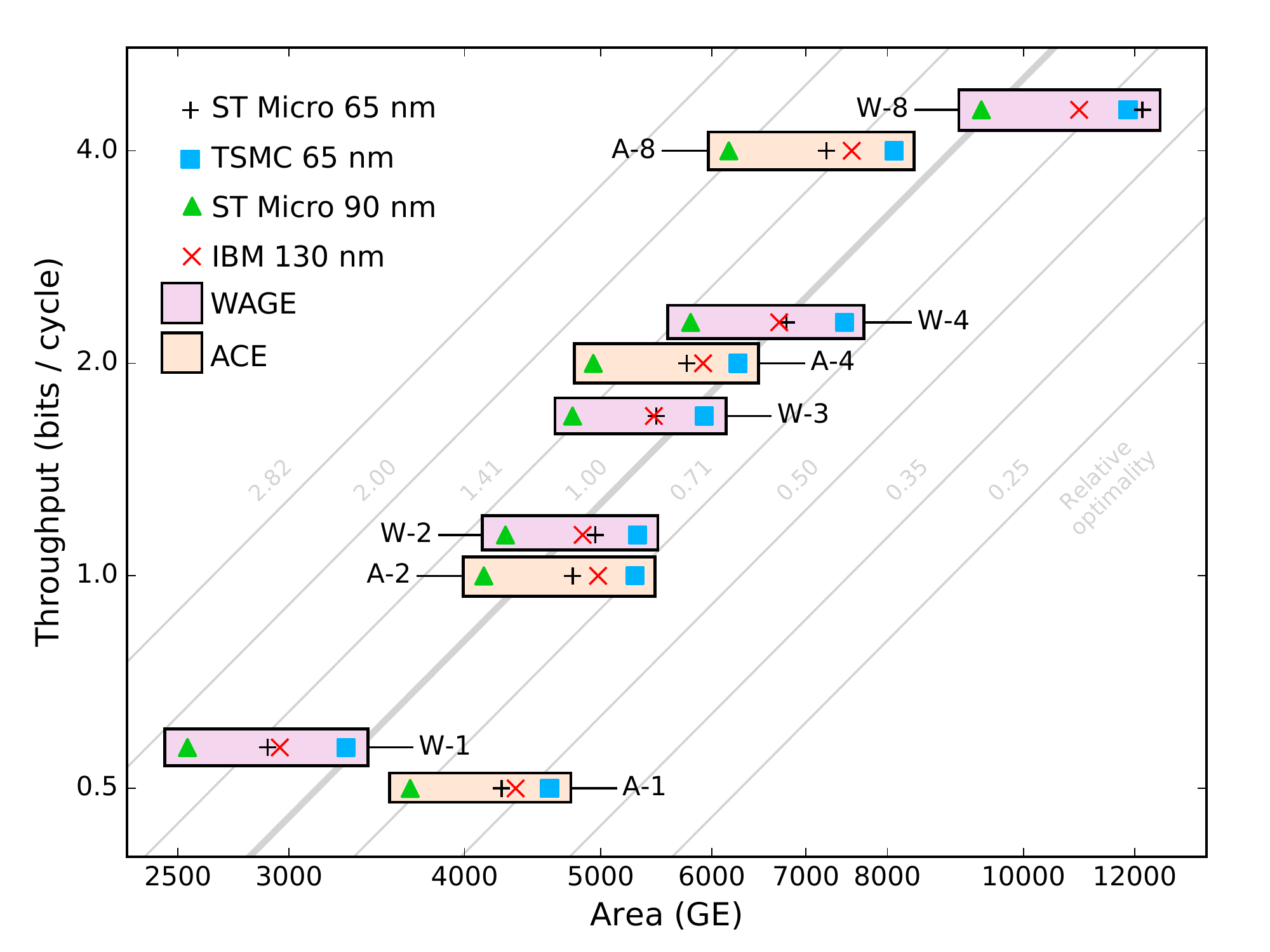}

        {\scriptsize
        \begin{tabular}{l}
        Throughput is measured in bits per clock cycle (bpc),
        and plotted on a log scale axis.\\  The area axis is
        scaled as log(Area$^2$).
          \end{tabular}
        }
        
	\caption{Area$^2$ vs Throughput}
	\label{fig.ta2graph}
\end{figure*}

\begin{table}[htbp]
  \footnotesize
  \begin{center}
    \captionof{table}{Post-PAR implementation results}
    \label{hwresults}
		\setlength{\extrarowheight}{0.3em}
		\begin{tabular}{|c||c|| c|c|c || c|c|c || c|c|c || c|c|c |}
\hline
& &\multicolumn{3}{c||}{ST Micro 65\,nm}  &\multicolumn{3}{c||}{ TSMC 65\,nm}  
& 	\multicolumn{3}{c||}{ST Micro 90\,nm} &\multicolumn{3}{c|}{IBM 130\,nm}	\\
\cline{3-14}
Label & Tput  & A & f 	& E & A & f 	& E 
& A & f 	& E & A & f 	& E\\
 \ [A/W-$p$] & [bpc]  &[GE] &  [MHz]	& [nJ] & [GE] & [MHz]	& [nJ]
& [GE] &  [MHz]	& [nJ] &[GE] &  [MHz]	& [nJ]\\

			\hline	
				\multicolumn{14}{|c|}{{\ace}  }	\\
				\hline
\textsf{A-1}&	0.5	
		& 4250	& 720	& 27.9	
		& 4600	& 705   & 20.1	
		& 3660	& 657	& 62.2 	
		& 4350	& 128	& 46.8	
\\ \hline
\textsf{A-2}&	1	
		& 4780	& 618	& 18.4 	
		& 5290	& 645	& 12.4	
		& 4130	& 628	& 35.8 	
		& 4980	& 88.9	& 29.4	
\\  \hline
\textsf{A-4}&	2	
		& 5760	& 394	& 15.1 	
		& 6260	& 588	& 8.51	
		& 4940	& 484	& 25.4 	
		& 5910	& 90.5	& 21.1	

\\  \hline
\textsf{A-8}&	4	
		& 7240	& 246	& 11.4 	
		& 8090	& 493	& 6.40	
		& 6170	& 336	& 19.4 	
		& 7550	& 63.2	& 18.4	
\\
\hline
				\multicolumn{14}{c}{}	\\[-1.5ex]
				\hline
								\multicolumn{14}{|c|}{{\wage}  }	\\
				\hline
\textsf{W-1}& 0.57 
		& 2900	& 907	& 20.0 	
		& 3290	& 1120	& 13.0	
		& 2540	& 940	& 39.2 	
		& 2960	& 153	& 30.4  
\\ \hline

\textsf{W-2}& 1.14
		& 4960	& 590	& 19.1 	
		& 5310	& 693	& 10.6	
		& 4280	& 493	& 34.4 	
		& 4850  & 98.5	& N/A\footnote{~} 
\\ \hline
\textsf{W-3}&	1.68
		& 5480	& 397	& 20.4 	
		& 5930	& 527	& 10.7	
		& 4770	& 414	& 31.2 	
		& 5460	& 79.6	& 26.5	
\\ \hline
\textsf{W-4}& 2.29
		& 6780	& 307	& 24.0 	
		& 7460	& 387	& 12.1	
		& 5790	& 277	& 32.9 	
		& 6700	& 51.9	& 33.4	
\\ \hline
\textsf{W-8}& 4.57
		& 12150	& 192	& 38.5 	
		& 11870	& 204	& 19.9	
		& 9330	& 137	& 49.9 	
		& 10960	& 34.5	& 59.9	
\\ \hline			
		\end{tabular}
		\end{center}

  Note: Energy results done with timing simulation at 10\,MHz.
  
\end{table}

Figure~\ref{fig.ta2graph} shows area$^2$ vs.\@ throughput for both
{\ace} and {\wage} with different degrees of parallelization, denoted
by W-$p$ and A-$p$ ($p=1,2,3,4,8$).  The throughput axis is scaled as
log(Tput) and the area axis is scaled as log\((\text{area}^2)\).  The
grey contour lines denote the relative optimality of the circuits
using Tput/area$^2$.  Throughput is increased by increasing the degree
of parallelization (unrolling), which reduces the number of clock
cycles per permutation round.

In comparing to the logic-synthesis with minimal optimization results
in Figure~\ref{fig:area-analysis}, the area for {\ace} has decreased
by 1.6\% and the area for {\wage}
has decreased by 2.7\%.
As parallelization is increased, {\wage}'s area
grows faster than \ace's.  For {\ace}, only the {\simecki} S-boxes and
a little bit of circuitry for the constant LFSR needs
to be replicated (approximately 500\,GE).   {\wage} requires the
feedback, WGPs, and small SB S-boxes be replicated.  The area increase
is larger than that of {\ace}, and also less consistent.
Going from
$p{=}1$ to $p{=}8$ results in $1.72\times$ area increase for {\ace}
and $3.80\times$ for {\wage} on average.  Optimality for {\wage} reaches a maximum at $p{=}3$.
For {\ace}, optimality continues to increase beyond $p{=}8$.

As can be seen by the relative constant size of the shaded rectangles
enclosing the data points, the relative area increase with
parallelization is relatively independent of implementation
technology.

Table~\ref{hwresults} represents the same data points as
Figure~\ref{fig.ta2graph} with the addition of maximum frequency (f,
MHz) and energy per bit (E, nJ). Energy is measured as the average value
while performing all cryptographic operations over 8192 bits of data
at 10 MHz. As the {\ace} throughput increases, energy per bit
decreases consistently, despite higher circuit area and, therefore,
power consumption. However, this is not the case with \wage.
When unrolling a combinational circuit, glitches increase
exponentially with circuit depth.  The critical path in {\ace}'s
{\simecki} S-box has only 2 gates, while the critical path in {\wage}'s
WGP contains 10 gates.  The significantly longer critical path,
combined with the larger overall area of WGP, cause energy consumption
to increase with parallelization for {\wage}.

Table~\ref{lwc-others} summarizes the area on ST Micro 65\,nm of the
LWC submissions\cite{lwcr1} that included synthesizable VHDL or
Verilog code. Table~\ref{lwc-others} reports the area results obtained
using the ST Micro 65\,nm process and tool flow from this paper and
the results reported in the submission.
The various ciphers use different protocols and
interfaces, sometimes provide different functionality (\eg, with or
without hashing), and use different key sizes.  As such, this analysis
is very imprecise, but gives a rough comparison to {\ace} and
\wage\ results.  As the LWC competition progresses and the hardware
API matures, more precise comparisons will become possible.  This
preliminary analysis indicates that {\ace} and {\wage} are among the
smaller cipher candidates.

\begin{table}[htbp]\footnotesize
  \centering
  \caption{Area of synthesizable LWC round 1 candidates on ST Micro 65\,nm (post-PAR)}
  \label{lwc-others}

  \begin{tabular}{|l l | c | c c |} \hline
  					&				 & This work &\multicolumn{2}{c|}{Reported in submission documents \cite{lwcr1}} \\
    \textbf{Cipher} & \textbf{Module} & \textbf{Area} (kGE)  & \textbf{Area} (kGE)
                    & ASIC technology used\\ \hline
    Drygascon & \verb|drygascon128_1round_cycle| & 29.6 &N/A &\\ \hline
    Gage & \verb|gage1h256c224r008AllParallel| & 10.4 &N/A &\\ \hline
    Lilliput-AE & \verb|lilliputaei128v1 encryptdecrypt| & 9.9 &4.2
                    &theoretical estimate for 5 lanes\\ \hline
    Remus & \verb|remus_top| &7.4		&3.6 &TSMC 65nm\\ \hline
    Subterranean & \verb|crypto_aead simple_axi4_lite| & 6.5 &5.7 &
                                                                     FreePDK
                                                                    45nm\\ \hline
    Triadx & \verb|triadx1| & 1.5 & & \\ \hline
    Thash & \verb|thash1| & 1.5 & & \\ \hline
  \end{tabular}
\end{table}

\section{Conclusion}

The goal of the {\ace} and {\wage} design process was to build on the well
studied Simeck S-Box and Welch-Gong permutation.  The overall
algorithms were designed to lend themselves to efficient
implementations in hardware and to scale well with increased
parallelism.  {\ace} has a larger internal state: 320 bits, vs 259 for
{\wage}, but the {\ace} permutation is smaller than that of {\wage}.
This means the non-parallel version of {\wage} is smaller than that of
{\ace}, but as parallelism increases, {\wage} eventually becomes
larger than {\ace}.  At 1 and 2 bits-per-cycle, the designs are
relatively similar in area.  A number of the NIST LWC candidate
ciphers provided synthesizable source code.  A preliminary comparison
with these ciphers on ST Micro 65\,nm indicates that {\ace} and
{\wage} are likely to be among the smaller candidates.

\textbf{Acknowledgements} This work benefited the collaborative
environment of the Communications Security (ComSec) Lab at the
University of Waterloo, and in particular discussions with {\ace} and
{\wage} crypto team of Riham
AlTawy, Guang Gong, Kalikinkar Mandal, and Raghvendra Rohit.

\def\c{\textsuperscript{1}}
\def\h{\textsuperscript{2}}

\end{document}